\documentstyle[epsfig, emulateapj5]{aastex}
\reversemarginpar
\makeatletter

\newenvironment{figurehere}
  {\def\@captype{figure}}
  {}
\makeatother

\begin{document}
\submitted{}
{\it To appear in the Proceedings of the STScI Symposium, "The Dark Universe: Matter, Energy, and Gravity" (April 2 - 5, 2001), ed. M. Livio}
\title{Simulating the X-ray Forest in a $\Lambda$CDM Universe}
\author{Taotao Fang, Greg L. Bryan\altaffilmark{1} AND Claude R.Canizares}
\affil{Department of Physics and Center for Space Research, Massachusetts Institute of Technology}
\affil{NE80-6081, 77 Massachusetts Avenue, Cambridge, MA 02139}
\altaffiltext{1}{Hubble Fellow}

\begin{abstract}

Numerical simulations predict that a large number of baryons reside in intergalactic space at temperatures between $10^{5}-10^{7}$ K. Highly-ionized metals, such as \ion{O}{7} and \ion{O}{8}, are good tracers of this ``{\it warm-hot intergalactic medium}'', or WHIM. For collisionally-ionized gas, the ionization fraction of each ion peaks  at some particular temperature (``peak temperatures''), so different ions can therefore trace the IGM at different temperatures. We performed a hydrodynamic simulation to study the metal distributions in the IGM. By studying the distribution functions of H- and He-like O, Si and Fe in a collisionally-ionized IGM and comparing with semi-analytic results based on the Press-Schechter formalism, we find: (1) ions with higher peak temperatures (for instance, \ion{Fe}{26}) tend to concentrate around virialized halos, which can be well described by the Press-Schechter distribution, ions with lower peak temperatures are found both in small halos (such as groups of galaxies) and in filaments; (2) lower peak temperature ions are more abundant and should be easily observed; (3) peculiar velocities contribute a significant part to the broadening of the resonant absorption lines.

\end{abstract}

\section{Introduction}

Cosmological N-body simulation plays an important role in studying cosmic structure formation and evolution. It is of particular interest to study the evolution of baryonic matter because it addresses an important question: why is there an apparent decrease in baryon density from high to low redshifts (see, e.g., \citealp{fhp98})? Numerical simulations indicated that the absence of baryons at low redshifts represents a yet undetected IGM: the ``{\it warm-hot intergalactic medium}'', or WHIM \citep{dco00}. For example, \citet{cos99} and \citet{dco00} found that the average temperature of baryons is an increasing function of time, with most of the baryons at the present time having a temperature in the range of $10^{5}-10^{7}$ K. Although invisible in the optical band, the highly ionized metals in WHIM gas might be detectable in UV and X-ray bands. 

Since highly-ionized metals, such as \ion{O}{7} and \ion{O}{8}, are good tracers of WHIM gas at $10^{5}-10^{7}$ K, in this paper we study the metal distributions in the IGM through hydrodynamic simulation. The idea is similar to that discussed in \citet{plo98}, \citet{hgm98} and \citet{fca00}: highly ionized metals would introduce absorption features in the X-ray spectrum of a distant quasar --- the ``{\it X-ray Forest}''. This paper has the following goals: (1) compare predictions of the X-ray forest from semi-analytic models to those from numerical simulations, this is important because each technique has its own strengths and weakness and by contrasting the two, we can determine which results are robust; (2) compute predictions for a wider range of ions, including \ion{O}{7}, \ion{O}{8}, \ion{Si}{13}, \ion{Si}{14}, \ion{Fe}{25}, \ion{Fe}{26}; (3) explore the nature of the gas that gives rise to the absorption, including its thermal state and its location in clusters, groups or filaments. 

\section{Numerical Simulation \label{sec:ns}}

We numerically simulated a cubic region of $100 h^{-1}$ Mpc, or 150 Mpc with $h=0.67$ on one side at $z ~\sim 0$. We model the dark matter with particles (as in a standard N-body code) but use the technique of {\it adaptive mesh refinement} (AMR) to follow the baryons. We refer to \citet{bry96} and \citet{nbr99} for further details. In this simulation the dark matter particle mass was $5 \times 10^{11}\ M_{\odot}$ and the smallest cell size (in the high density regions) was $49h^{-1}$ kpc. To generate spectra, we created a uniform grid which contained $512 \times 512 \times 512$ cells, with each cell having a size of $\sim 195h^{-1}$ Kpc, or $19.5\ km\ s^{-1}$ in velocity space. The simulation itself is a flat low-density cold dark matter model ($\Lambda$CDM) with $\Omega_{0} = 0.3$, $\Omega_{\Lambda} = 0.7$ and $\Omega_{b} = 0.04$. The amplitude of the initial power spectrum was fixed so that the {\it rms} overdensity in a spherical tophat with radius $8h^{-1}$ Mpc, $\sigma_8$, is $0.9$. We use a scale-invariant power spectrum with a primordial index of $n=1$. The initial cosmological redshift is $z=30$.

Little is known about the metallicity of the IGM. The metal abundance could be as low as $10^{-3}Z_{\odot}$ in the low column density Ly$\alpha$ forest systems \citep{lss98} or as high as $1Z_{\odot}$ in the interstellar medium (ISM) of some early-type, X-ray luminous galaxies \citep{mom00}. Since we will compare numerical simulation with the semi-analytic model in our previous paper \citep{fca00}, we assume uniform metalicity and use the metallicities observed from the intracluster medium, e.g., $0.5Z_{\odot}$ for oxygen, silicon and $0.3Z_{\odot}$ for iron \citep{mla96}. To investigate X-ray absorption in the spectrum of a background source, we need to understand the distribution of ionization states of metals in the IGM. To obtain the ion density, we assume the gas is in collisional equilibrium and we adopt the ionization fractions from \citet{mmc98}. We select helium- and hydrogen-like oxygen, silicon and iron. The peak temperatures of these ion species range from  $10^{5}$ K to $10^{9}$ K, which probe different temperature regions of the IGM. 

We have obtained the projected column density, $N^{i}$, distributions of all the six ions. Figure~\ref{f1} (for \ion{O}{7}) and Figure~\ref{f2} (for \ion{Fe}{26}) show two examples. In each figure, the ion column density distributions are plotted in four different panels with $N^{i} > 10^{12}$, $10^{13}$, $10^{14}$ and $10^{15}\ cm^{-2}$, respectively. The color bar on top of each panel indicates the logarithm scale from the minimum to the maximum column densities in each panel. 

\begin{figurehere}
\centerline{\psfig{file=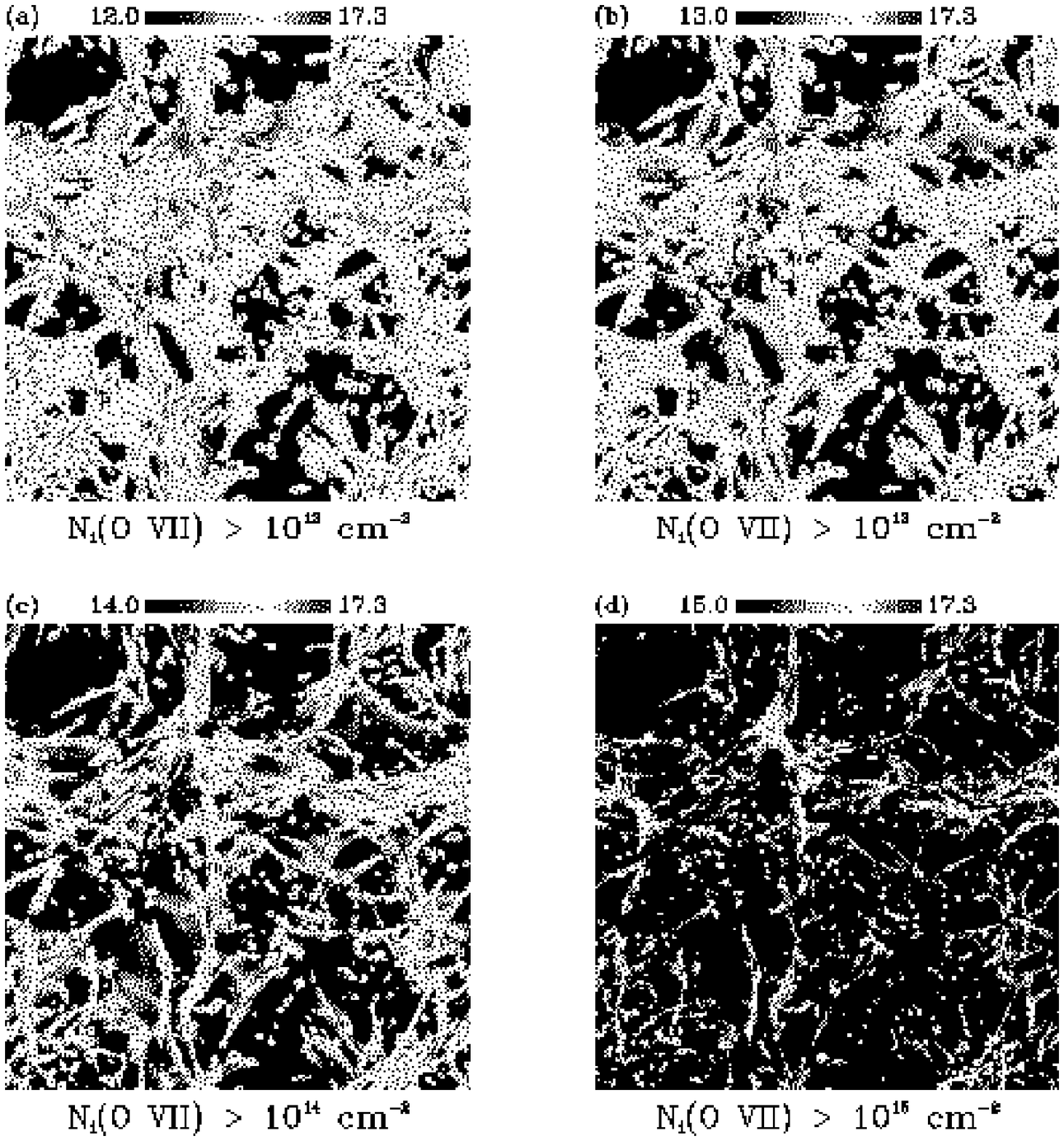,width=7cm}}
\caption{The projected density (column density, $N^{i}$) distributions of \ion{O}{7}, plotted in four different panels with $N^{i} > 10^{12}$, $10^{13}$, $10^{14}$ and $10^{15}\ cm^{-2}$, respectively. The color bar on top of each panel indicates the logarithm scale from the minimum to the maximum column densities in each panel. \label{f1}}
\end{figurehere}
\vspace{0.2cm}

By examining the column density distributions of all the six ions, we find that the distributions of these ions are closely related to the density and temperature distributions. Regions with higher baryon densities also have higher metal densities. Due to the fact that different ions have different peak temperatures, there are differences among the distributions of these ion species. For instance, many of the filaments that connect high density regions are shock-heated to temperatures above $10^{5}$ K. Since \ion{O}{7} has a peak temperature between $3\times10^{5}$ K and $2\times10^{6}$, we find that a large amount of \ion{O}{7} exists in filaments with a column density even higher than $10^{15}\ cm^{-2}$ (Figure~\ref{f1}d). On the other hand, \ion{Fe}{26} has a peak temperature of over $10^{8}$ K. In Figure~\ref{f2} we find that almost no \ion{Fe}{26} exists in filaments, and it all concentrates around those collapsed intersections. By comparing these two extremes, we find that generally \ion{Fe}{26} follows the distribution of virialized objects while most of \ion{O}{7} exists in filaments, and the distributions of other ion species vary between those of \ion{Fe}{26} and \ion{O}{7}.

\section{Spectral Analysis \label{sec:sa}}

The methodology we adopt is straightforward: a random line-of-sight (LOS) is drawn across the simulated region; temperature, baryon density and velocity are obtained at each point along the LOS; the ion density is calculated based on the metallicity and ionization fraction; the synthesized spectrum is then obtained from the optical depth along the LOS; finally, by fitting the resonant absorption lines in the ``{\it simulated}'' spectrum we obtain ion column densities ($N^{i}$) and the Doppler parameters ($b$). This method is similar to the method used in the Ly$\alpha$ forest; we refer to \citet{zan97} for a complete discussion and formalism. We draw a total of 5000 random LOS across the simulated region. For each LOS we generate a simulated spectrum and fit it by the ``{\it threshold-based}'' method. We then calculate the so-called ``X-ray forest'' distribution function (XFDF), $\partial^{2}P/\partial N^{i}\partial z$, defined as the number of absorption systems along the line of sight with a column density between $N^{i}$ and $N^{i}+dN^{i}$ per unit redshift.

\begin{figurehere}
\centerline{\psfig{file=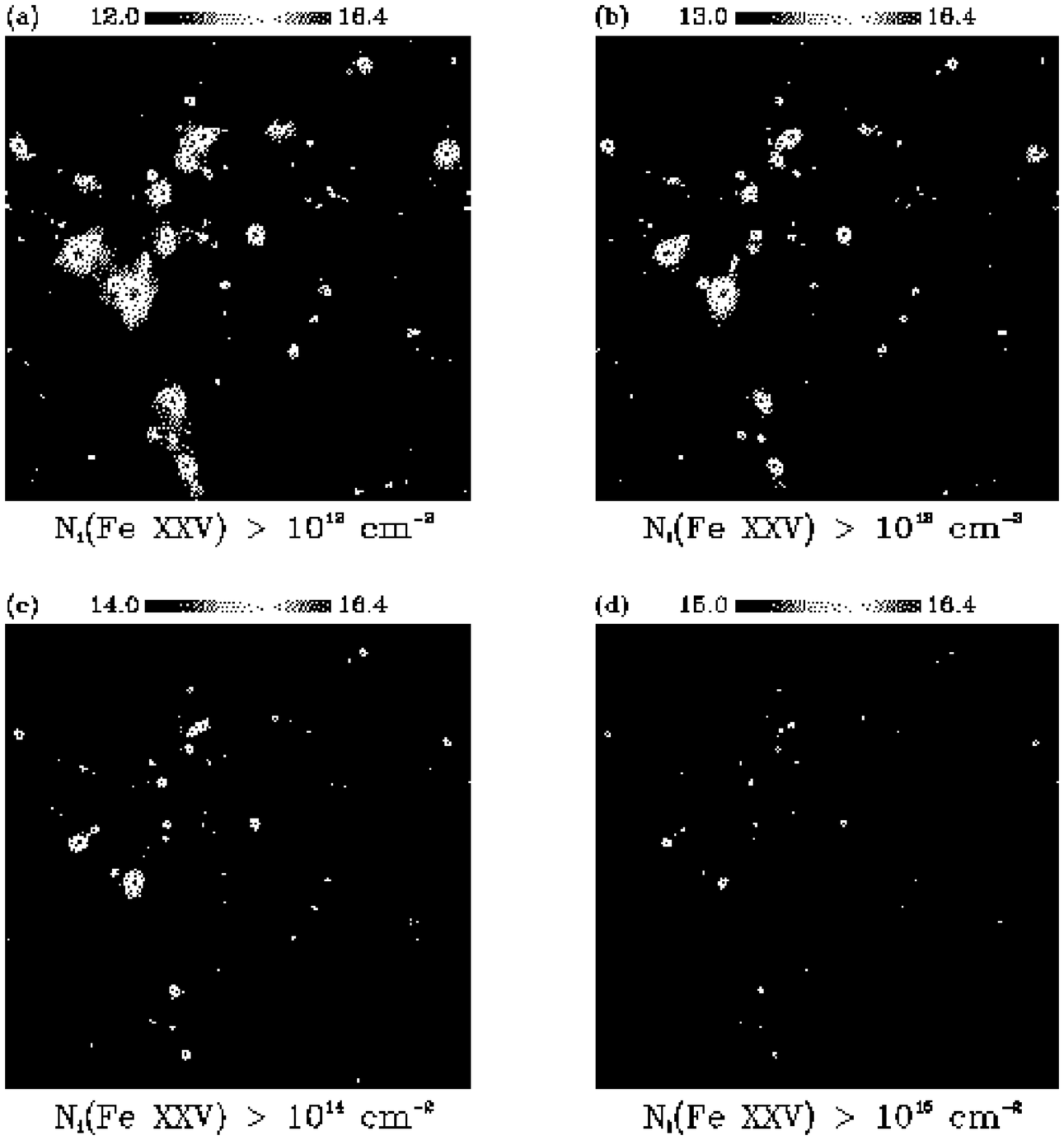,width=7cm}}
\caption{The projected column density distributions of \ion{Fe}{26}. \label{f2}}
\end{figurehere}
\vspace{0.2cm}

Based on the formalism in \citet{fca00}, we compare our numerical simulation with the analytic model. The Press-Schechter function \citep{psc74} predicts the spatial distribution of virialized objects. We would expect that the analytic model and the numerical simulation should give consistent results for the distribution of the high-peak-temperature ions, such as \ion{Fe}{25} and \ion{Fe}{26}, because these ions follow the distribution of virialized objects; on the other hand, we would not expect the low-peak-temperature ions, such as \ion{O}{7} and \ion{O}{8}, would have the same behavior because most of these ions are distributed in filaments.

Figure~\ref{f3} shows the XFDF predicted from the Press-Schechter approach for all the six ions (solid line in each plot) and also displays the distributions from the numerical simulation (the filled-dotted line). The most striking feature is that the predicted distributions fit the simulation very well at high column density regions, specifically at $N_{i} > 10^{16}\ cm^{-2}$ for both oxygen species and at  $N_{i} > 10^{15}\ cm^{-2}$ for silicon species and \ion{Fe}{25}. For \ion{Fe}{26}, both distributions agree very well in the whole column density range ($10^{12} - 10^{16}\ cm^{-2}$). This result can be well understood in the framework we describe at the beginning of this section: high-peak-temperature ions, such as \ion{Fe}{26}, follow the distribution of collapsed, virialized halos well described by the PS formalism. For low-peak-temperature ions, the high column densities also come from virialized structures, while those with low column density are distributed in those filaments which can not be described by the PS distribution. Thus except for \ion{Fe}{26}, at low column density the PS distributions are systematically lower than in the numerical simulation.

\begin{figurehere}
\centerline{\psfig{file=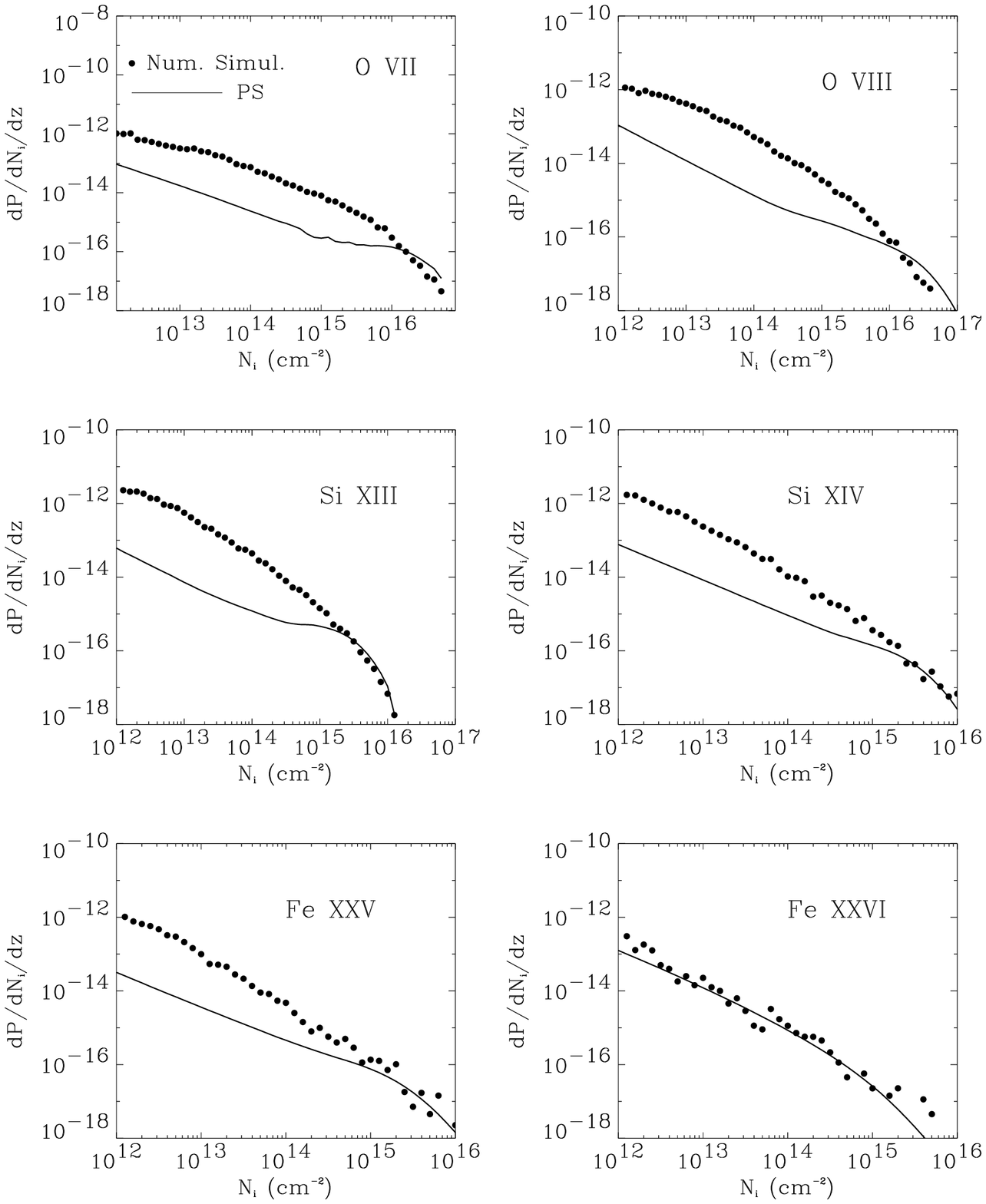,width=9cm}}
\caption{The XFDF predicted from the Press-Schechter approach (solid lines) and numerical simulation (filled-dotted lines). \label{f3}}
\end{figurehere}
\vspace{0.2cm}

Another feature seen in this figure is that all curves display a sharp cutoff somewhere between $10^{16}$ and $10^{17}\ cm^{-2}$. The reason for this apparent drop at high column density end is two fold: first, systems with high column density generally correspond to high temperature regions, and at high temperature the ionization fraction drops quickly and so does the ion column density; second, high column density systems are closely related to virialized objects, such as groups and clusters of galaxies, and the Press-Schechter theory predicts that the spatial density of virialized objects decreases exponentially as their masses (and temperatures) increase.

Doppler $b$-parameter governs the line width and can be affected by several factors, such as gas temperature (thermal motion), the Hubble flow and the  velocity structure. Figure~\ref{f4} shows the $b$-parameter distribution function for six ions (the filled circles). All six ions exhibit a low cutoff to the Doppler parameter. The cutoffs are primarily due to the cutoff temperatures for each ionization. 

To understand the role of different line broadening mechanisms, we first fit the $b$-parameter distribution with a Gaussian. We find that there exists a non-Gaussian tail in each plot. This tail can be well fitted by a lognormal distribution (the solid line): $f \propto \exp \left[-\log^{2}(b/\bar{b}_{ln})/2\sigma_{ln}^{2}\right]$ Here $\bar{b}_{ln}$ and $\sigma_{ln}$ are the mean velocity and the standard deviation of this lognormal distribution. We find this distribution naturally cuts off at low $b$, and also provide an excellent fit to the long tail. This non-Gaussian tail implies that the Hubble flow and peculiar velocity contribute a significant part to those ions with high $b$ values.
 
\begin{figurehere}
\centerline{\psfig{file=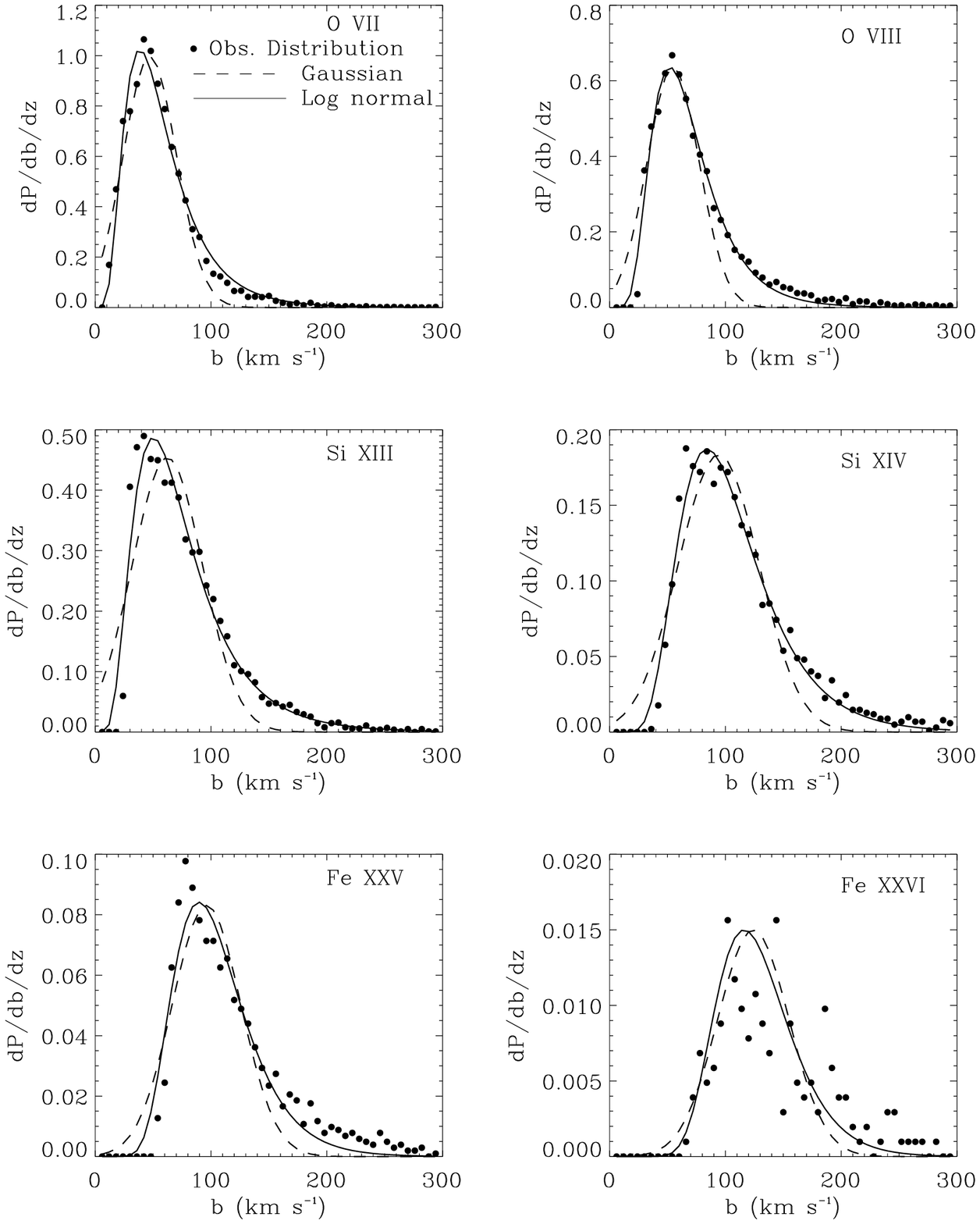,width=9cm}}
\caption{The $b$-parameter distribution function for six ions (the dotted lines), fitted with Gaussian distributions (the dashed lines) and log normal distributions (the solid lines). \label{f4}}
\end{figurehere}
\vspace{0.2cm}

\section{Summary and Discussion \label{sec:sd}}

In this paper we have performed a hydrodynamic simulation based on a $\Lambda$CDM model. By drawing random LOS and synthesizing the X-ray spectrum for each LOS, we obtained the XFDFs for hydrogen and helium-like O, Si, Fe. We have investigated the column density distributions of highly-ionized metals by comparing the numerical simulation to the analytic method described in \citet{fca00}. We also discuss the distribution of Doppler $b$-parameters. Our main conclusions can be summarized as follows:

\begin{enumerate}

\item Ions with low peak temperatures such as \ion{O}{7} and \ion{O}{8} trace small halos (galaxy groups) and filaments, while high peak ions such as \ion{Fe}{25} and \ion{Fe}{26} are only observed in the largest objects (galaxy clusters). For all ions, the strongest signal come from virialized halos; however, filaments show up as lower column density absorbers.

\item The low peak temperature ions are more abundant than ions with high peak temperatures. This can be understood from numerical simulations that most of the baryons have temperatures ranging from $10^{5}-10^{7}$ K and the baryon fraction decreases as the temperature increases. This implies that from the point view of observation, ions with low peak temperatures such as \ion{O}{7} and \ion{O}{8} should be more easily observed.

\item We compared the column density distribution of absorbers predicted by the simulations against an analytic model developed in \citet{fca00} and found that the numerical simulation confirms our predictions: ions with high peak temperatures ($T > 10^{8}$K) follow the distributions of collapsed halos, which can be described by the Press-Schechter distribution; for ions with low peak temperatures, those lines with high column density come from virialized structures, while those with low column density are distributed in diffuse filaments which are not well described by the PS distribution.

\item Higher excitation ions produce lines which are generally wider (as determined by the Doppler $b$-parameter) than low excitation lines, although there is a wide distribution in all cases. This distribution is well fit by a log normal distribution. The sharp lower cutoff of this distribution is provided by the fact that there is a minimum temperature required to create the ion in the first place. Peculiar velocities in the gas also contribute significantly, providing a non-Gaussian tail at the high-$b$ end.

\end{enumerate}

The largest uncertainty in spectral synthesis comes from the metallicity: we did not include any gas processes in our numerical simulation. Although these gas processes are not important in heating and cooling gas with $T>10^{5}$ K, the IGM metals enriched by these processes are crucial in UV/X-ray observations: it is these metals which produces detectable features. We assume a uniform metallicity in generating the synthesized spectrum, however, we would expect that metallicity should be determined by the metal enrichment history of the IGM, which in turn depends on a variety of sources: the UV/X-ray background radiation, supernova input, etc, and these are likely to be different in low and high density regions. It is important to incorporate all these processes to obtain a self-consistent result of metal distributions and ionization structures in the IGM. This is the goal of our future numerical simulations.

\acknowledgments{TF thanks the support from MIT/CXC team. This work is supported in part by contracts NAS 8-38249 and SAO SV1-61010. Support for GLB was provided by NASA through Hubble Fellowship grant HF-01104.01-98A from the Space Telescope Science Institute, which is operated by the Association of Universities for Research in Astronomy, Inc., under NASA contract NAS 6-26555.}

\end{document}